\documentclass[12pt]{article} 
\usepackage{amssymb,amsmath,amsthm,graphicx,pst-all,url,textcomp}
\title{Algorithms using Java for Spreadsheet Dependent Cell Recomputation\footnote{This work was supported by US Army CECOM Contract DAAB07-01-C-C201, and
SPAWAR PMW 176-1.  \textcopyright 2002 The MITRE Corporation.  All rights reserved. }}
\author{Joe Francoeur \\ The MITRE Corporation \\ 202 Burlington Road \\ Bedford, MA 01730-1420\\ jfrancoe@mitre.org}

\newcommand{\Cellfields}[2]{%
\psframebox{%
  \begin{tabular}{c}
    \textnormal{#1}\\
    \scriptsize{#2}
  \end{tabular}}}

\newtheorem{prp_valform}{Property}
\newtheorem{prp_nocircref}[prp_valform]{Property}
\newtheorem{thm}{Theorem}
\newtheorem{alldeps}[thm]{Theorem}

\begin{document}
\nocite{*} 
\maketitle
\begin{abstract}
Java implementations of algorithms used by spreadsheets to automatically recompute the set of cells dependent on a changed cell are described using a mathematical model for spreadsheets based on graph theory.  These solutions comprise part of a Java API that allows a client application to read, modify, and maintain spreadsheet data without using the spreadsheet application program that produced it.  Features of the Java language that successfully improve the running time performance of the algorithms are also described.  
\end{abstract}
\bibliographystyle{is-plain}

\section{Introduction}

This paper describes algorithms for the recomputation of spreadsheet cells.  The assumed context for such a recomputation occurs when a cell's value is changed. In general, a cell is dependent on several others for its value as defined by its formula.  Thus, to maintain the integrity of the spreadsheet, the reading of a cell value requires the recomputation of this cell once any of the cells on which it depends has changed.

The algorithms of this paper form the basis of \emph{ExcelComp} \cite{local:ExcelComp}, a Java \cite{java:1_3} application program interface (API) written by the author that allows the client application to read a specially formatted Microsoft Excel \cite{Excel:homepage} (henceforth referred to as ``Excel'') spreadsheet output file, and then make changes to cell values within the ExcelComp representation of this spreadsheet.  Changes to cell values are followed by the automatic recomputation of dependent cell values using ExcelComp methods.  ExcelComp thus allows the client programmer to provide its users with both the data and behavior of an existing spreadsheet without the use of the original spreadsheet application program that produced it.

During the development of ExcelComp, it was realized that choices of algorithms to perform cell recomputation involve two principal trade-offs: 1) ease of use, and 2) running time performance.  On the one hand, one mode of ExcelComp can simply load a file at run time that represents the spreadsheet, and then provide its services.  While this mode is satisfactory for many tasks, it is unsuitable for those that require a large number of cell recomputations to support dynamic updates to real-time outputs, for example, the updating of an on-screen map that depends on thousands of cell recomputations.  To support this latter task, a second mode was developed that allows faster cell recomputation at the expense of a less convenient installation procedure for the spreadsheet representation.

These considerations make ExcelComp an efficient, platform- and vendor-independent Java API that provides built-in spreadsheet emulation for application end-users.  In particular, end-users are relieved of the burden of conducting their spreadsheet tasks outside the domain of their running application.  In addition to the efficiency won by executing spreadsheet tasks natively, ExcelComp also obviates the need for costly additional licenses required for multiple users of the application software that produced the spreadsheet.  Being written in the modern Java programming language allows the client programmer to easily integrate ExcelComp's functionality into current software development efforts.

While other descriptions of spreadsheet algorithms are available \cite{msdn:ExcelRecalc}\cite{gnu:gnumeric-dependencies}, this paper is distinctive in its use of graph theory to improve the reader's ability to visualize the algorithms, and to provide a basis for a proof of algorithm correctness.  It also presents solutions that leverage features in the object-oriented Java API that lead to succinct, yet powerful code.     

This paper focuses on the subject algorithms and the specific features of the Java language used by ExcelComp that are well-suited for their implementation.  Readers interested in a more detailed specification of ExcelComp from the client programmer's perspective may contact the author.

\section{A Scenario}

Before getting into the technical details that comprise this report, it would be helpful to consider a motivational scenario.

Consider a spreadsheet of financial data, where subtotals, interest earned, and a grand total might be some examples of computed quantities that each depend on entries in several cells.  Analysts may use such a spreadsheet to play ``what-if'' games by varying values in cells that will affect some target cell, such as interest earned.  The spreadsheet program would then automatically recompute all cells that are dependent on the ones changed.  This capability of a spreadsheet is its hallmark, and distinguishes it from a simple table of values that have no computational relationship to one another.

Suppose that a computer program needs this spreadsheet of information and auto-update capability to carry out its tasks.  This program is to provide its users with the what-if capability, and therefore requires not only cell values, but also cell formulas.  Since it needs to emulate the recomputation function of the spreadsheet, it must implement algorithms that return the same recomputed values as the spreadsheet program.  It is these algorithms of dependent cell recomputation that are the subject of this report.

\section{Modes}

ExcelComp has two modes of operation:
\begin{description}
\item[Interpreted mode] requires the reading of an eXtensible Markup Language (XML) \cite{xml:spec} representation of a spreadsheet.  Once this file is parsed and loaded into ExcelComp's data structures, the subject algorithms are implemented via ExcelComp methods.  It is called \emph{interpreted}, because cell formulas are interpreted at run time using a custom parser that recognizes a subset of Excel's formula language.
\item[Compiled mode] uses cell-specific Java classes, created as an offline preprocessing task, to evaluate a cell by recursively evaluating each child cell referenced in its formula's parse tree.
\end{description}

The interpreted mode is the slowest of the two.  It has the advantage, however, of requiring less preprocessing, namely just the creation of the XML input file.  During the development of ExcelComp, this XML file was produced by running an Excel macro \cite{local:XMLmacro}.  It is also easier in this mode for the client programmer to provide the application user the flexibility to apply ExcelComp to a different spreadsheet by simply changing a filename reference in ExcelComp's constructor.  The comprehensive update of all dependent cell values upon the change of a constant cell in this mode allows the ExcelComp user to highlight the newly computed dependents.  Such an application was developed by the author, where changed cells are shown in a JTable \cite{java:JTable} with changed values highlighted in red to allow the user to gain insight into the impact of the change of a cell value.

The compiled mode is \emph{much} faster than the interpreted mode, and should be preferred in cases where a client demands exceptionally high execution time performance, e.g., providing a real-time screen update that depends on the recomputations.  The preprocessing needed for compiled mode includes the generation of Java source code that implements the formulas of the spreadsheet.  This source code generation was automated by using a Java class that uses classes produced by parser generators \cite{JLex,CUP} to implement a custom parser for a subset of the Excel formula language.  In general, Java classes that represent each cell formula in the spreadsheet must be created, and then be referenced by the classpath option for the Java virtual machine (JVM) \cite{java:JVM}.

\section{Computation Model}

To provide a lingua franca for the discussion to follow, we need to identify parts of a spreadsheet that are useful to us.  While the intent is to have a model that is generic, platform-, and vendor-independent, the use of Excel as a reference implementation for ExcelComp influenced the latter's design.  The language of this paper will be similarly influenced, however it is germane to any spreadsheet that adheres to the computation model described here.  While a queue-based computation model has been successfully developed \cite{msdn:ExcelRecalc}, we will find it advantageous to develop a model using graph theory.  In particular, proof of correctness of the algorithms can benefit from such a treatment.

There are many ways to present data using a spreadsheet.  For example, two principal classes of representation provided by Excel are the \emph{workbook}, and the \emph{chart}.  We consider only the tabular computation environment found in a workbook.  The term \emph{spreadsheet} will thus be used as a synonym for workbook.

A \textbf{spreadsheet} is a finite set of cells arranged as a matrix.  A \textbf{cell} is a set that contains three elements of interest: 
\begin{enumerate}
\item a value, 
\item a formula, and 
\item a cell reference.
\end{enumerate}

A cell's \textbf{value} is the result of the computation specified by its formula.  In general, this value may be a real number, a string, or some other data type.  To simplify this model, we will assume that these values are real.  A cell's \textbf{formula} is an expression that defines a cell's value as a function, $f$, of a subset of the spreadsheet's cell values.  Let $\mathcal{C}$ denote the set of cell values for some spreadsheet.  More formally then, we have 

\begin{equation} \label{formula}
f:\mathcal{C}^n \to \mathcal{C}, 
\end{equation}
where $\mathcal{C}^n$ is the $n$-fold Cartesian product of $\mathcal{C}$ for some positive integer $n$.  For this model, we define $\mathcal{C}=\mathbb{R}$.  In general, a formula expresses a composite of functions in the form \eqref{formula}.  A formula that is not composite has constant values for its arguments.  Such a formula is termed a \textbf{constant formula}.

A \textbf{cell reference} is an ordered pair that specifies a cell uniquely within the spreadsheet.  The Excel ``A1 reference style''\cite{Excel} will be used, where the first element specifies the column, and the second element specifies the row.  For example, B3 designates the cell at the intersection of the second column and the third row.  We use Xi to denote a variable whose value is a cell reference.  In the context of a formula, a cell reference is mapped to its corresponding cell value according to \eqref{formula}.  We thus see that, in general, a cell reference refers to a composite function that is defined by the formula for that cell.  For example, if cell M1 depends on N1 and P2, and N1 depends on Q3, and P2 depends on R1, then the value of M1 expressed as a composite function is M1(N1(Q3),P2(R1)).  For this expression to be fully resolved, the formulas for both Q3 and R1 must be constant formulas.

To successfully develop the subject algorithms, the scope of the set of spreadsheets to be considered must be defined.  This will be done by identifying properties that serve as axioms for the spreadsheets of interest.  Spreadsheets that satisfy the stated properties are termed \textbf{admissible}.

There is a cohesive relationship between a cell's formula and its value, as described in the following property.

\begin{prp_valform} \label{prp:valform}
A cell's value is completely determined recursively by its formula.
\end{prp_valform}

To be clear, Property~\ref{prp:valform} states that a cell's value depends only on its formula, but that formula in general depends on other cell values that are, in turn, dependent only on their formulas.  This recursion ends when a cell with a constant formula is reached, thus resolving all of the recursive cell value references.

The formulas of a spreadsheet define a potentially involved relation among its cells.  Given cell Xi, its formula may be a constant formula, or a non-constant formula that defines Xi's value as a function of other cell values in the spreadsheet.  In the latter case, we say that Xi is \textbf{dependent} on the cells referenced in its formula.  That is, the \emph{value} of Xi depends on the \emph{values} of the cells referenced in its formula.  The term ``value'' will often be omitted when the context of ``dependent'' is clear.  We also refer to each cell referenced in Xi's formula as a \textbf{child} of Xi.  Similarly, Xi is a \textbf{parent} of its children.  A parent and any parent of a parent is termed an \textbf{ancestor}.  A child and any child of a child is termed a \textbf{descendant}. 

\begin{figure}[!ht]
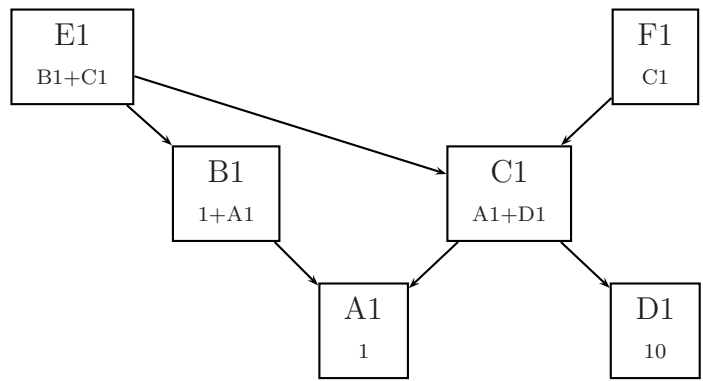

\begin{center}

\psmatrix[mnode=r,colsep=.5,rowsep=.5,nodesep=3pt]
  \Cellfields{E1}{B1+C1} &  &  &  & \Cellfields{F1}{C1} \\
     & \Cellfields{B1}{1+A1} &  & \Cellfields{C1}{A1+D1} &  \\
     &  & \Cellfields{A1}{1} &  & \Cellfields{D1}{10}
\endpsmatrix

\ncline{->}{1,1}{2,2}
\ncline{->}{2,2}{3,3}
\ncline{->}{2,4}{3,3}
\ncline{->}{1,5}{2,4}
\ncline{->}{2,4}{3,5}
\ncline{->}{1,1}{2,4}

\end{center}
\caption{Example of a Spreadsheet Dependency Graph} \label{fig:cellrel}
\end{figure}

The relation among the dependent cells in a spreadsheet may be represented as a weakly connected directed graph, $G(V,E)$, where $V$ is the set of vertices, and $E$ is the set of edges.  Figure~\ref{fig:cellrel} illustrates this \textbf{spreadsheet dependency graph} with an example that will be used throughout this paper.

Each vertex of the graph is a cell represented as a box.  The cell reference is given at the top of each cell box, and its corresponding formula is given in smaller type at the bottom of the box.  It is understood that the value of a cell is assigned the value computed by its formula.  The cell values are omitted from the cell boxes in Figure~\ref{fig:cellrel} for brevity.  Note that C1 has two parents: E1 and F1.  This lack of a unique parent in general for each cell precludes regarding this structure as a rooted tree.  Although the more general \emph{graph} is the appropriate data structure for representing cell dependencies in a spreadsheet, we will find that rooted trees will also be useful in the algorithms to be described.

In describing Figure~\ref{fig:cellrel}, some of the basics of graph theory, using \cite{Di} as a guide, will be described as needed.  

In Figure~\ref{fig:cellrel}, the set of vertices $V$ are the cells, and the set of directed edges $E$ is defined according to the parent/child relationships.  Each edge is directed from a parent to a child.  The set $E$ is a subset of the ordered pairs of $V$.  Let a sequence of vertices be ordered such that $v_{i-1}$ is a parent of $v_i$, and let $v_{i-1}v_i$ denote the edge directed from $v_{i-1}$ to $v_i$.  A sequence of edges and vertices that has the form $\{v_0v_1,v_1v_2,\ldots,v_{n-1}v_n\}$ for distinct $v_i$ is defined as a \textbf{directed path} linking $v_0$ and $v_n$.  A directed path is a \textbf{directed cycle} if it consists of 2 or more vertices, and $v_n=v_0$.

We are now led to an important stipulation concerning spreadsheets.

\begin{prp_nocircref} \label{prp:nocircref}
An admissible spreadsheet contains no directed cycles.
\end{prp_nocircref}

The Excel term for directed cycle is \textbf{circular reference}.  Property~\ref{prp:nocircref} thus states that circular references are prohibited.

A graph from the subset of graphs just described is termed a \textbf{directed acyclic graph} or \textbf{dag} \cite[section B.4]{CLR}.

Readers familiar with the GNU Make tool \cite{gnu:make} will recognize this dependent cell recomputation problem as being analogous to the problem solved by Make: automatic determination of the pieces of code that require recompilation, and issuing the appropriate commands to bring the program up to date.  Make uses a dependency graph model.  See \cite{RecursiveMake} for details and illustrations of Make's dependency graphs, including a description of pitfalls concerning the proper use of Make to ensure correct dependency graph construction in large projects.

\section{Interpreted mode}

This section describes those algorithms that are implemented in the interpreted mode of ExcelComp.  The integrity of the spreadsheet is preserved in this mode by recomputing all dependent cells of a cell whose constant formula (value) has changed.  This behavior ensures that, upon the commitment of a new value to a cell, the entire spreadsheet will be updated to reflect the change.  This matches the default behavior of Excel, where it is termed \emph{automatic calculation}.  Pseudocode is given in this section to highlight the salient features of the ExcelComp interpreted mode algorithms; the actual code differs in some of the implementation details.

\subsection{Dependency Set Generation}

Suppose that we are examining a spreadsheet for the first time, and have no a priori knowledge of its contents.  Say we want to modify cell A1.  By this, we mean that A1 has a constant formula that is to be changed to another constant formula.  The more general act of modifying or adding a non-constant formula will not be discussed here; it is assumed that non-constant formulas remain fixed throughout our analysis.

Consider the impact that this change has on the cells that are dependent on A1 in Figure~\ref{fig:cellrel}.  First, this change in the formula causes a recomputation of A1's value.  This change will, in general, affect the values of all cells whose formulas reference A1.  These cells, B1 and C1, are \textbf{directly dependent} on A1, and will need to be recomputed as a result.  In general, the values of these direct dependents will change as a result of recomputation.  These direct dependents must then be considered in the same light as A1; that is, we need to find and recompute the direct dependents of the direct dependents of A1.  These cells are E1 and F1.  From the point of view of A1, these latter cells are \textbf{indirect dependents} of A1.

The algorithm for discovering the set of dependent cells of a given cell is thus recursive.  Let $d$ be a set-valued function $d:2^\mathcal{C} \to 2^\mathcal{C}$ that computes the set of direct dependents of a subset of cells from $\mathcal{C}$.  (Here, $2^\mathcal{C}$ denotes the set of all subsets of $\mathcal{C}$, also known as the \emph{power set}.)  The procedure just described can now be expressed as

\begin{equation}\label{rrel}
A_{i+1}=d(A_i),
\qquad A_i \in 2^\mathcal{C},
\qquad A_0=\{\mathrm{A1}\}.
\end{equation}

$A_0$ is set to $\{\mathrm{A1}\}$ in \eqref{rrel} to reflect Figure~\ref{fig:cellrel}, but in general it will be assigned to the set containing the cells that were changed.

Dependency Set Generation may be recognized as an implementation of the \emph{breadth-first search} (BFS) algorithm for graphs described in \cite{CLR}.  The ``frontier between discovered and undiscovered vertices'' described in \cite{CLR} applied here divides two generations of dependencies, i.e., child/parent, parent/grandparent, etc.  Also, note that we begin with a child, and then discover ancestors, in reverse to the naming convention used in \cite{CLR}.

For recurrence relation \eqref{rrel} to be practical, we must be assured that it terminates.  Indeed, an essential property of an algorithm is its finiteness; according to \cite{Knuth},``An algorithm must always terminate after a finite number of steps.''  This assurance is given now as a theorem.

\begin{thm} \label{thm:rrelends}
Recurrence relation \eqref{rrel} terminates.
\end{thm}
\begin{proof}
Because the indices of $A_i$ in the recursion are strictly increasing, it is sufficient to show that $\exists i_{max} \ni i \le i_{max}$.  

Assume that the spreadsheet under consideration has $N$ cells.  Let $\|A_i\|$ denote the number of parents of $A_{i-1}$.  Because of Property~\ref{prp:nocircref}, the number of candidate parents for $A_0$ is $N-1$, since a cell cannot be a parent of itself (thereby creating a circular reference).  Similarly, the number of available parents for $A_1$ is at most $N-2$, since both the children and grandchildren of $A_2$ must be excluded to avoid circular references.  In general then, 
\begin{equation*}
\|A_i\|\le N-i,
\end{equation*}
 and in particular, $\|A_{N}\|=0$.  At this point, no parents are available to continue further.  We have thus shown that $i \leq N$; that is, $i_{max}=N$.
\end{proof}

The final product of Dependency Set Generation is formed by taking the union of the sets of dependent cells found in \eqref{rrel}.  Assuming that the final index computed in \eqref{rrel} is $n$, and letting $\mathcal{D}$ be the set of dependent cells, we have

\begin{equation}\label{union}
\mathcal{D}=\bigcup_{i=1}^n A_i,
\qquad A_i \in 2^\mathcal{C}
\end{equation}

\begin{figure}[!ht]
\begin{center}
\newbox\tmpboxa 
\setbox\tmpboxa=\vbox{\advance\hsize by-2\fboxrule\advance\hsize by-2\fboxsep 
\begin{verbatim}
Input: Set of cells for which dependent cells will be found
Output: Set of cells that are dependent on the input
        set of cells
DepSetGen(depSet,m)
{
  initial_size = depSet.size();
  for (k=m through initial_size-1;k++) {
    for (j=0 through SPRSHEET_SIZE-1;j++) {
    // Find all direct dependents of depSet[k].
      if (sprsheet[j].formula contains depSet[k].ref) {
        depSet.add(sprsheet[j].ref);
      }
    }
    DepSetGen(depSet,initial_size);
  }
  if (m == 0) {
    depSet.delete(depSet[0]);
  }
}
\end{verbatim}}
\noindent\fbox{\box\tmpboxa} 
\caption{ Dependency Set Generation Algorithm} \label{alg:depsetgen}
\end{center} 
\end{figure}

\subsubsection{Example}

The Dependency Set Generation algorithm is codified in Figure~\ref{alg:depsetgen}.  Let us apply this algorithm to finding all dependents of A1 in Figure~\ref{fig:cellrel}.

Assume the number of cells in the spreadsheet, \verb|SPRSHEET_SIZE|, is 6.  The array \verb|sprsheet| holds all the cells in the spreadsheet.  Each element of \verb|sprsheet| has the fields \verb|ref| and \verb|formula| for cell reference and formula, respectively.  The array \verb|depSet| will be built up to contain the cell references of all of the dependents of its initial value. 

We make the initial call to \verb|DepSetGen| with \verb|depSet| initialized to contain A1, and the \verb|depSet| element marker \verb|m| set to 0.  This marker's value is the index of the cell in array \verb|depSet| whose dependents are sought.  Variable \verb|initial_size| is set to the number of elements in \verb|depSet|.  Since \verb|depSet| contains only A1, \verb|initial_size| = 1.  Loop counter \verb|k| ranges from 0 through 0.  The inner loop checks to see whether any spreadsheet cell formula contains A1.  If it does, the cell reference is added to \verb|depSet|.  At the time where the inner loop is finished, both B1 and C1 are appended to \verb|depSet|.  At the bottom of the outer loop, it is time to make the first recursive call to \verb|DepSetGen|.

The actual parameters passed to \verb|DepSetGen| are the newly updated 3-element \verb|depSet|, and the initial size of \verb|depSet| before the loops, 1.  Now entering the first recursive call of \verb|DepSetGen|, \verb|m| is 1, \verb|initial_size| is 3, \verb|depSet| consists of A1, B1, and C1.  Loop variable \verb|k| ranges from 1 through 2.  The first time through the inner loop finds all dependents of B1, and adds the one dependent found, E1, to \verb|depSet|.  When \verb|k| is 2, the inner loop finds all dependents of C1, and adds the one dependent found, F1, to \verb|depSet|.

Upon the next recursive call, the marker is set to the next unexamined element of \verb|depSet|, E1 at index 3.  No dependents are found.  Similarly for F1 at index 4.  Finally the outer loop is skipped, and control is eventually returned to the original call of \verb|DepSetGen|, where \verb|m| is 0.  Lastly, the \verb|if| statement is executed, and the initial element A1 is removed from \verb|depSet|, since A1 is not dependent on itself.  It is assumed \verb|depSet| contains just one cell during the initial call to \verb|DepSetGen|.  Though not shown in Figure~\ref{alg:depsetgen}, an additional step of removing duplicate cell references from $\mathcal{D}$ is required to ensure that all of its elements are unique.

To conclude this section, it will be shown that the Dependency Set Generation algorithm just described indeed finds all dependents of a given cell.

\begin{alldeps}
The Dependency Set Generation algorithm identifies all dependents of its input set of cells.
\end{alldeps}
\begin{proof}
The proof is by contradiction.  Assume we have a spreadsheet with cells $C_i$, $i=0,1,2,\ldots,N-1$.  Suppose $\exists C_k \in \mathcal{C}$ that is dependent on $C_0$, but was not identified by the Dependency Set Generation algorithm.  Then by Property~\ref{prp:valform}, this dependence of $C_k$ on $C_0$ must be due only to $C_k$'s recursive formula.  There must then be a directed path in $C_0$'s dependency graph from $C_k$ to $C_0$. Since $C_k$ was not identified, there is at least one cell $C_j$ in this path that was not found during the recursion.  But this contradicts the step in the algorithm that says to find \emph{all} direct dependents of $C_{j-1}$.
\end{proof}

\subsection{Recomputation of Dependency Set Members}

Once the dependency set $\mathcal{D}$ has been generated, the process of recomputing these cells can begin.

To evaluate a cell, designated the \textbf{original cell}, each argument in its formula, the \textbf{original formula}, must be evaluated.  This evaluation is in general a recursive procedure.  By drawing a directed edge from the original cell to each cell referenced in the formula (one edge per cell in the formula), we get a rooted tree that is rooted at the original cell.  By applying this algorithm recursively on each cell in the formula, we get several paths, each of which ends at a leaf having a constant-formula cell.  The resulting rooted \textbf{call tree} is just a subset of the spreadsheet dependency graph.  We may then start at the leaves of the tree to construct a string of formulas of the cells in a given path in the direction back toward the root.  Each string is a self-contained sequence of formulas that allows the original cell to be evaluated.  

Lastly, once all the arguments in the original formula have been evaluated, the original cell can be evaluated.  Those arguments in any formula that are not members of $\mathcal{D}$ need not be recomputed; their current values can be used instead.

The foregoing procedure is an implementation of the \emph{depth-first search} (DFS) algorithm described in \cite{CLR}.  In this case, the use of the terms ``predecessor'' and ``descendant'' in \cite{CLR} is consistent with our usage.  However, we do not use ``timestamping.''

We can see a lot in common here with the Dependency Set Generation algorithm.  Once again, we see a recursive procedure being described, although it is not as easily expressible in one line as in \eqref{rrel}.  Instead, we will codify the algorithm in the pseudocode given in Figure~\ref{alg:depseteval}.

\begin{figure}[!ht]
\begin{center}
\newbox\tmpboxa 
\setbox\tmpboxa=\vbox{\advance\hsize by-2\fboxrule\advance\hsize by-2\fboxsep 
\begin{verbatim}
Input: Original Cell 
Output: Original Cell with newly computed value
depth = 0;  // Initial value (global)
EvalCell(cell)
{
  for (i=0 through cell.nchildren-1;i++) {
     depth++;
     EvalCell(cell.child[i]);
  }
  parser_str.append(cell.ref + '=' + cell.formula + ';');
  if (depth == 0) {
    cell.value = parse(parser_str);
  }
  depth--;
  return cell;
}
\end{verbatim}}
\noindent\fbox{\box\tmpboxa} 
\caption{Dependency Set Evaluation Algorithm} \label{alg:depseteval}
\end{center}
\end{figure}

\subsubsection{Example (continued)}

Continuing the example begun in the Dependency Set Generation section, consider again the dependency graph in Figure~\ref{fig:cellrel}.  Suppose that E1 is to be evaluated.  The well-known \emph{left-to-right, post-order tree traversal} algorithm will be used to specify the order of evaluation of E1's descendants.  During the first call to \verb|EvalCell|, the recursion depth variable \verb|depth| is initialized to 0.  In addition to the \verb|ref| and \verb|formula| fields, assume that a \verb|cell| object also contains a field \verb|nchildren| that gives the number of children in its \verb|formula| field.  The object \verb|cell| also contains a \verb|child| array, each of whose elements is a \verb|cell| representing each child in its formula.  The elements of \verb|child| are stored in order of occurrence in its \verb|formula|, element 0 being the leftmost child, and element \verb|nchildren-1| being the rightmost.

We thus begin by calling \verb|EvalCell| with \verb|cell.ref| set to E1.  The recursive evaluation of E1's call tree begins with the leftmost cell reference in E1's formula, B1.  $\mathrm{B1} \in \mathcal{D}$, and therefore must be recomputed.  We then begin with the leftmost element of its formula, and find that it is the constant 1.  This is a constant, and thus requires no further evaluation; we move to the next element, A1.  A1 is the changed cell, and its value is known, thus no further analysis is needed.

We have now reached the end of B1's formula, allowing us to compute its value.  We thus go back up to E1 to process the next argument in its formula, C1. $\mathrm{C1} \in \mathcal{D}$, and therefore must be recomputed.  The leftmost child of C1's formula is A1, and has already been evaluated.  C1's next child, $\mathrm{D1} \notin \mathcal{D}$, and thus does not have to be recomputed.  Its current value of 10 is used.

We have now recursively evaluated all of the children of E1's formula, and completed the building of \verb|parser_str|.  This string may then be passed to a parser for evaluation of E1.  ExcelComp uses an LALR parser developed by the author using the tools JLex \cite{JLex} and CUP \cite{CUP}.  LALR stands for LookAhead Left-to-right identifying the Rightmost production, and is described in \cite{Grune}.  For this example, the string is:

\begin{center}
\texttt{A1=2;B1=1+A1;A1=2;D1=10;C1=A1+D1;E1=B1+C1;.}
\end{center}

Here, ``\verb|=|'' stands for assignment, and ``\verb|;|'' delimits each assignment.  It is assumed that the parser stores values via the assignment statements, and that these values may be retrieved at points later in the parse string.  The history of the construction of the parser string is summarized in Table~\ref{tbl:parse}.

\begin{table}[!ht]
\begin{center}
\begin{tabular}{|l|l|l|l|}
\hline
\textsf{depth} & \textsf{cell.ref} & \textsf{i} & \textsf{parser\_str} \\
\hline \hline
0 & E1 & 0 & \\
\hline
1 & B1 & 0 & \\
\hline
2 & A1 & 0 & A1=2; \\
\hline
1 & B1 & 0 & A1=2;B1=1+A1; \\
\hline
0 & E1 & 1 & A1=2;B1=1+A1; \\
\hline
1 & C1 & 0 & A1=2;B1=1+A1; \\
\hline
2 & A1 & 0 & A1=2;B1=1+A1;A1=2; \\
\hline
1 & C1 & 1 & A1=2;B1=1+A1;A1=2; \\
\hline
2 & D1 & 0 & A1=2;B1=1+A1;A1=2;D1=10; \\
\hline
1 & C1 & 1 & A1=2;B1=1+A1;A1=2;D1=10;C1=A1+D1; \\
\hline
0 & E1 & 1 & A1=2;B1=1+A1;A1=2;D1=10;C1=A1+D1; \\
  &    &   & E1=B1+C1; \\
\hline
\end{tabular}
\end{center}
\caption{History of Parser String Construction for E1} \label{tbl:parse}
\end{table}

Note that there is a redundant assignment ``\texttt{A1=2}'' in \verb|parser_str|.  This is an example suggesting efficiency enhancements that can be seen to improve the performance of these algorithms.  Improvements include, but are not necessarily limited to:
\begin{enumerate}
\item Recompute a cell value only once.  
\item Do not recursively evaluate cells that are not in $\mathcal{D}$. 
\end{enumerate}

Although these improvements surely are desirable for minimizing the number of algorithm steps, experience with their use in \mbox{ExcelComp} revealed that the time to run the additional code required to implement these improvements largely cancels out the benefits of fewer cell evaluations.

More ideas concerning the speed-up of interpreted mode are given in the Algorithm Complexity section.

\section{Compiled mode}

While the compiled mode agrees with interpreted mode with respect to the preservation of spreadsheet integrity, it uses a clever postponement of computation technique to update a dependent cell's value just prior to the reading of its value via its accessor method.  Using this deferred recomputation strategy, the execution time associated with recomputing those dependent cells whose values are never accessed is eliminated.  This deferred recomputation is similar to Excel's \emph{manual calculation} mode, where the user specifies when recomputation is to occur thus deferring immediate recomputation.  It is also similar to the \emph{mark-sweep} garbage collection algorithm \cite{gc:survey}.  

Compiled mode is implemented by ExcelComp's use of the \verb|Cell| API \cite{local:Cell}.  The highlight of this mode is preserving spreadsheet integrity while improving the running time performance of ExcelComp.  Several techniques are used to meet this goal, including the use of:
\begin{enumerate}
\item the Java Reflection API
\item Hash containers, and
\item Deferred recomputation of dependent cells.
\end{enumerate}

\subsection{The Cell Class}

\verb|Cell| is an abstract Java base class that provides a framework for modeling the cells of a spreadsheet, each of which is represented by a class derived from \verb|Cell| named \verb|CellXi|, where \verb|Xi| denotes the A1-style reference to its corresponding cell.

On its initial invocation, \verb|Cell|'s accessor class method \verb|getCell| instantiates a \verb|CellXi| object via the Java Reflection API \cite{java:Reflection}.  This technique allows a Java class to instantiate another class whose name is created at run time by the calling method.  To improve the efficiency of subsequent accesses, \verb|Cell| has a class variable \verb|workbook| to reference a \verb|HashMap| of references to previously instantiated \verb|CellXi| objects.  Similarly, each \verb|CellXi| object has a \verb|HashSet| named \verb|dependencies| that contains references to \verb|CellXi| objects that correspond to cells that are direct dependents (parents) of the \verb|CellXi| object that owns \verb|dependencies|.

In a \verb|CellXi|'s constructor, each child's instance method \verb|addDependency| is called to add Xi to that child's \verb|dependencies| set.  The use of \verb|getCell| during this process causes each child's object, on its first access, to be initialized with its constructor.  A DFS traversal of \verb|CellXi|'s call tree thus occurs so that each traversed parent appears in each of its children's \verb|dependencies| set.  The set of all such parents is the set of \textbf{discovered ancestors}.  Both \verb|workbook| and \verb|dependencies| contain only minimal subsets of the full set of their respective data that describes the entire spreadsheet as determined by the history of \verb|Cell| method calls.  These subsets are updated as necessary, and suffice for computing correct results when recomputation of cell values is necessary.

\subsection{Preprocessing}

\begin{figure}
\includegraphics[scale=0.70]{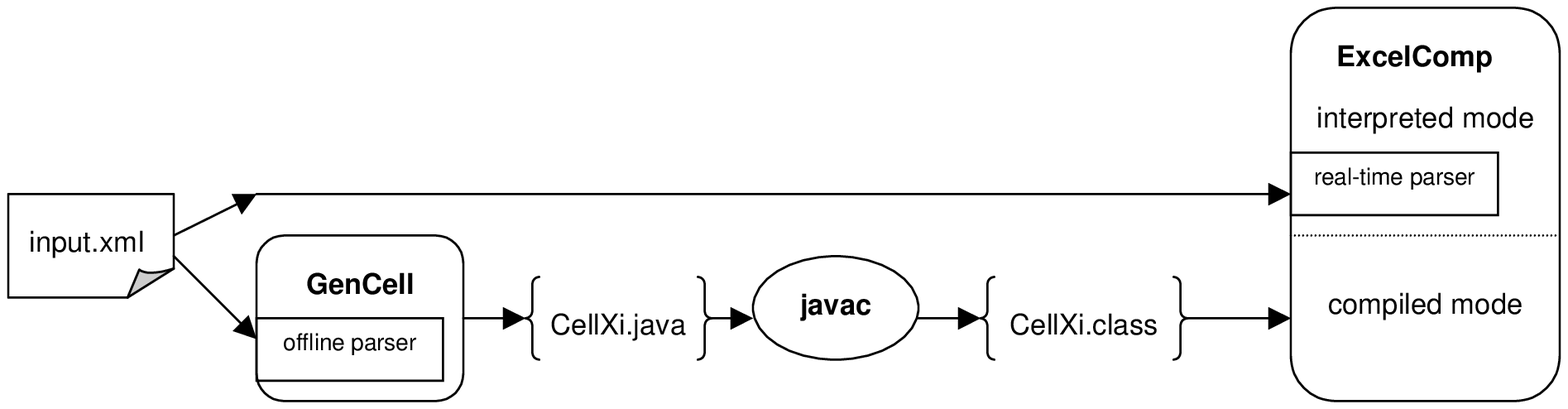}
\caption{Preprocessing requirements for ExcelComp modes} \label{fig:modes}
\end{figure}

Figure~\ref{fig:modes} details the differences in the preprocessing requirements for the two modes of ExcelComp.

The preprocessing necessary for compiled mode begins with the same XML input file used in interpreted mode.  A Java API named \verb|GenCell| \cite{local:GenCell} is run on this input file to produce a set of Java source files.  Each source file defines a \verb|CellXi| class that corresponds to a spreadsheet cell.  \verb|GenCell| uses a JLex/CUP-based parser similar to that used in interpreted mode to allow the translation of a supported subset of the Excel formula language into the appropriate Java statements.  The important difference between the two parsers is in their output.  In interpreted mode, the ExcelComp parser returns a string at run time that represents a newly-computed value.  For ExcelComp's compiled mode, the \verb|GenCell| parser returns Java source code that is written to a set of source files as an offline preprocessing task.  Once \verb|GenCell| has completed generating all of the source files, the Java compiler is run to compile these files into class files of executable bytecode.  The ExcelComp user must then ensure that the Java classpath contains the appropriate references to allow the client application to find the classes at run time.

\subsection{Dependency Set Generation}

The nature of the preprocessing performed in compiled mode allows ExcelComp to handle Dependency Set Generation as a distributed, on-demand task rather than as an explicit set of steps conducted immediately after a cell value is changed as in interpreted mode.

\subsection{Deferred Recomputation of Dependent Cells}

Rather than requiring that the immediate recomputation of all members of $\mathcal{D}$ occur after a cell's value has been changed, compiled mode defers the recomputation of any member of $\mathcal{D}$ until the client programmer requests its value via an ExcelComp accessor method.  This algorithm saves considerable time when compared to its interpreted mode counterpart, seeing that the recomputation of many dependencies whose value is never sought is avoided.

\subsection{Getting a Value}

Each \verb|CellXi| object's \verb|getValue| method contains the Java encoding of the formula for cell Xi.  The value of the cell is recomputed and stored as the instance variable \verb|val| only when that object's boolean \verb|dirty| flag is true; this flag thus allows this method to avoid unnecessary recomputation.  When recomputation is unnecessary, \verb|getValue| just returns the value of \verb|val|.

\subsection{Setting a Value}

When a \verb|CellXi| object's \verb|setValue| method is called, its instance variable \verb|val| is set to the desired value, and a DFS traversal of all of Xi's discovered ancestors is performed to ensure that each such ancestor's \verb|dirty| flag is set to \verb|true|.  This ensures that all ancestors' values are recomputed on a subsequent call to an ancestor's \verb|getValue| method.  Note that recomputations are only done if an ancestor's \verb|getValue| method is called, thus saving the time of recomputing ancestor values that may never be accessed.

\subsection{Example}

The following example illustrates the workings of compiled mode using Figure~\ref{fig:cellrel}.  Given the spreadsheet represented by this graph, we will use compiled mode to set A1's value to 2, and then get the value of E1.  Observe that E1's value with A1 set to 1 is 13.  By changing A1's value to 2, we expect the new value of E1 to be 15.

The first statement given by the client program is: 
\begin{quote}
\verb|Cell.getCell("A1").setValue(2);|
\end{quote}
The initial part of this statement, \verb|Cell.getCell("A1")|, creates a new instance of \verb|CellA1|, since one does not yet exist.  This new instance is created using the \verb|forName| class method in the package \verb|java.lang.Class| \cite{java:Class}.  \verb|CellA1|'s base class constructor \verb|Cell()| is first called to set \verb|CellA1|'s \verb|dirty| flag to \verb|true|.  \verb|CellA1|'s \verb|val| variable is set to 1 by its constructor.  The last part of this statement calls \verb|Cell|'s \verb|setValue| method to set \verb|val| to 2.  In general, \verb|setValue| recursively marks all of A1's \emph{discovered} ancestors as dirty.  However, since no ancestors have yet been discovered, and thus no corresponding \verb|CellXi| objects have yet been instantiated, no such marking occurs here.  We are thus left with one instance of \verb|CellA1| that is marked as dirty and has a value of 2.

We now get the value of E1.  The appropriate statement is:
\begin{quote}
\verb|Cell.getCell("E1").getValue();|
\end{quote}
The first part of the statement behaves in the same way as for \verb|A1| described above, only now the newly created object is an instance of \verb|CellE1|.  In addition, \verb|CellE1|'s constructor initiates a DFS traversal of all of E1's descendants to update each descendant's \verb|HashSet dependencies|.  For this example, both \verb|CellB1|'s and \verb|CellC1|'s \verb|dependencies| sets are updated by having a reference to \verb|CellE1| added.  Note that \verb|CellC1|'s \verb|dependencies| set does not refer to \verb|CellF1| since, although F1 is dependent on C1, it is not a descendant of E1.  Such a reference to \verb|CellF1| need only be added if F1 is the subject of future method calls.  The last part of this statement checks to see whether the instance of \verb|CellE1| has been marked as dirty.  Since \verb|CellE1| was just constructed anew, it is marked as dirty and thus its value must be recomputed (in this case, computed for the first time).  The statement in \verb|CellE1|'s \verb|getValue| method that accomplishes this is:
\begin{quote}
\begin{verbatim}
this.val = Cell.getCell("B1").getValue() + 
           Cell.getCell("C1").getValue();
\end{verbatim}
\end{quote}
Since the objects for B1, C1, and D1 were all marked as dirty during the DFS traversal in \verb|CellE1|'s constructor, each object's \verb|getValue| method will recompute the value for that cell.  Each object's \verb|getValue| method resets that object's \verb|dirty| flag to \verb|false| after the recomputation.  Subsequent accesses to \verb|CellXi| values that have not been affected by a change to a descendant's value simply return the value stored in \verb|val| with no recomputation necessary.

\section{Performance}

\subsection{Tests}

ExcelComp was tested for its running time performance in the SATCOM Availability Analyst (SA2) Java application \cite{SA2}.  The addition of ExcelComp API calls to SA2's map display function was chosen for test due to its demanding requirement that 8,518 data points be updated for an on-screen Mercator map in such a way that the user is not burdened by long wait times for a complete update of the map.  Processing each of the data points required 3 calls to ExcelComp methods; 2 of these calls each changed a cell value from the input spreadsheet, and the last call read back a cell value of interest from the newly updated spreadsheet.  The map display function was selected and run 10+ times in each mode to characterize ExcelComp's performance.  Running times associated with the first invocation of the map function were greater than subsequent trials, and thus were considered outliers and removed from the representative data.  These larger values probably reflect JVM-related setup steps that are not required on subsequent trials.

The tests were conducted on a Hewlett-Packard HP OmniBook 4150 B running under Microsoft Windows 98 on a Pentium III 650 MHz processor.  SA2 was run using the Sun Microsystems JVM version 1.3.  Running times were computed as the difference in the start and end times returned by the Java method \verb|System.currentTimeMillis()| \cite{java:System}.

\begin{table}[!ht]
\begin{center}
\begin{tabular}{|l|l|l|}
\hline
  & \textsf{Compiled Mode} & \textsf{Interpreted Mode} \\
\hline \hline
\textsf{Average} & 677 & 347057 \\
\hline
\textsf{Sample Standard Deviation} & 116 & 254 \\
\hline
\end{tabular}
\end{center}
\caption{ExcelComp Running Time (milliseconds) Performance over 10 Samples} \label{tbl:performance}
\end{table}

The results of the performance tests are summarized in Table~\ref{tbl:performance}.  The sample standard deviation is computed as the positive square root of the \emph{unbiased} sample variance.  The large difference in performance between the modes highlights how compiled mode can provide a very acceptable performance level in a case where interpreted mode, requiring over 5 minutes to complete, would be unacceptably slow.  In this particular case, it is essential that compiled mode be chosen to make the use of ExcelComp feasible.

\subsection{Algorithm Complexity}

The graph traversal algorithms that underlie ExcelComp are well known to be efficient.  Both DFS and BFS have running times that are linear in the size of the graph's adjacency list.  Specifically, BFS is $O(V+E)$, and DFS is $\Theta(V+E)$ \cite[sect.~22.2, 22.3]{CLR}.

Interpreted mode does not construct an adjacency list, and could very well benefit from a redesign to create this list upon the loading of the spreadsheet in the ExcelComp constructor.  Because this construction must take place at run time, the user would incur a one-time performance penalty for this initialization step.  The absence of an adjacency list suggests that interpreted mode's running time is probably greater than the linear time cited above.

Compiled mode, on the other hand, does use a variation of adjacency lists in \verb|CellXi|'s \verb|dependencies| set.  However, while an adjacency list stores references to children, \verb|dependencies| stores references to parents of \verb|CellXi|.  Compiled mode incurs a setup penalty during the discovery of cells, but subsequent accesses to \verb|CellXi| objects are more efficient through the use of \verb|dependencies| and \verb|workbook|.  Its use of adjacency lists, DFS traversals, and the efficient Java collections framework suggests that compiled mode has a running time that is close to $\Theta(V+E)$.

\section{Conclusion}

We have seen that a graph representation of spreadsheet cell dependencies provides insight into the requirements of the algorithms used for the automatic recomputation of dependent cells.  Straightforward implementation of well-known graph traversal algorithms suffices for correct recomputation, however the adaptation of a well-studied garbage collection algorithm along with facilities made available in the Java language enable client programs to run much faster, given some additional preprocessing.

The client programmer should choose the mode of ExcelComp according to an analysis of the application's run time requirements and the tradeoffs between the modes as described in this paper.

\section{Acknowledgements}

I would like to thank Deborah Schuh and Dr. Joseph Rushanan for their reviews and helpful comments on the contents of this paper.

\bibliography{ExcelComp-Algorithms}

\end{document}